\documentclass[conference]{IEEEtran}
\IEEEoverridecommandlockouts
\usepackage{cite}
\usepackage{amsmath,amssymb,amsfonts}
\usepackage{algorithmic}
\usepackage{graphicx}
\usepackage{textcomp}
\usepackage{xcolor}

\usepackage[latin1]{inputenc}
\usepackage{colortbl}
\usepackage{soul}
\usepackage{multirow}
\usepackage{pifont}
\usepackage{color}
\usepackage{alltt}
\usepackage[hidelinks]{hyperref}
\usepackage{enumerate}
\usepackage{siunitx}
\usepackage{breakurl}
\usepackage{epstopdf}
\usepackage{pbox}
\usepackage{booktabs}
\usepackage{eurosym}
\usepackage{authblk}

\def\BibTeX{{\rm B\kern-.05em{\sc i\kern-.025em b}\kern-.08em
    T\kern-.1667em\lower.7ex\hbox{E}\kern-.125emX}}

\usepackage{eso-pic}
\usepackage{url}
\AddToShipoutPictureBG*{
  \AtPageUpperLeft{%
    \put(0,-40){\raisebox{15pt}{\makebox[\paperwidth]{\begin{minipage}{21cm}\centering
      \textcolor{gray}{This article has been accepted for publication in the proceedings of the 2021 54th IEEE International Symposium on Circuits \\and Systems (ISCAS). Content may change prior to final publication. DOI: 10.1109/ISCAS51556.2021.9401362} 
    \end{minipage}}}}%
  }
  \AtPageLowerLeft{%
    \raisebox{25pt}{\makebox[\paperwidth]{\begin{minipage}{21cm}\centering
      \textcolor{gray}{ \copyright 2021 IEEE.  Personal use of this material is permitted.  Permission from IEEE must be obtained for all other uses, in any current or future media, including reprinting/republishing this material for advertising or promotional purposes, creating new collective works, for resale or redistribution to servers or lists, or reuse of any copyrighted component of this work in other works.
      }
    \end{minipage}}}%
  }
}

\begin{document}

\title{H-Watch: An Open, Connected Platform for AI-Enhanced COVID19 Infection Symptoms Monitoring and Contact Tracing
}

\author[1]{
Tommaso Polonelli
}
\author[1]{
Lukas Schulthess
}
\author[2]{
Philipp Mayer
}
\author[1]{
Michele Magno
}
\author[2]{
Luca Benini
}
\affil[1]{
Project Based Learning, ETH Zurich, Zurich, Switzerland
}
\affil[2]{
Integrated Systems Laboratory, ETH Zurich, Zurich, Switzerland
}

\maketitle

\begin{abstract}
The novel COVID-19 disease has been declared a pandemic event. Early detection of infection symptoms and contact tracing are playing a vital role in containing COVID-19 spread. As demonstrated by recent literature, multi-sensor and connected wearable devices might enable symptom detection and help tracing contacts, while also acquiring useful epidemiological information.
This paper presents the design and implementation of a fully open-source wearable platform called H-Watch. It has been designed to include several sensors for COVID-19 early detection, multi-radio for wireless transmission and tracking, a microcontroller for processing data on-board, and finally, an energy harvester to extend the battery lifetime. Experimental results demonstrated only 5.9~mW of average power consumption, leading to a lifetime of 9 days on a small watch battery. Finally, all the hardware and the software, including a machine learning on MCU toolkit, are provided open-source, allowing the research community to build and use the H-Watch.  
\end{abstract}


\section{Introduction}
COVID-19 has been declared a pandemic by the World Health Organization (WHO) and poses a significant challenge for healthcare infrastructure around the world. Continuous vital sign monitoring for symptoms of severe pneumonia and sepsis, such as blood~\cite{bedford2020covid} oxygen saturation level (SpO2~$<$93\%), respiratory rate ($>$30 breaths/minute), heart rate, body temperature, fatigue, coughing detection, and blood pressure can assist in the early recognition of high-risk patients~\cite{percivalle2020prevalence}. For example, saturation values below 95\% are a symptom of hypoxemia (reduction in the presence of oxygen in the blood), usually due to a decrease in gas exchange at the pulmonary alveoli level, a typical symptom of the worsening of some viral pneumonia~\cite{grifoni2020targets}. Continuously monitoring those parameters and tracking users in their movements is crucial not only to early detect infections but also to follow its diffusion~\cite{zens2020app}. However, continuous patient monitoring and tracking are ultra-challenging for many key issues, such as privacy~\cite{cho2020contact}, complexity of the monitoring, early recognition~\cite{zens2020app}, long-term operation with mobile battery-operated devices, and discontinuous connectivity, among others. 

%

Thanks to technology advancements in low-power integrated circuits (ICs), sensors, and wireless protocols have enabled the practical and daily use of light-weighted and unobtrusive wearable devices~\cite{dang2019novel}, where electronics are worn on the human body or hidden into clothes. Among other products, smartwatches are a massive commercial reality with hundreds of products specifically designed for fitness and entertainment applications. As a containment measure for the COVID-19 pandemic, back-tracing contacts and personal interactions via smartphones is becoming increasingly important. Its active contribution in identifying potential outbreaks has been demonstrated~\cite{ashraf2020smartphone}. While GPS tracking, in addition to non-negligible privacy and security implications~\cite{ashraf2020smartphone}, poses serious limits to adequate indoor coverage, the use of BLE technology is considered as a good compromise~\cite{oosterlinck2017bluetooth} between the tracing precision (in a range from 10~cm to 10~m) and user privacy~\cite{essa2019improve}; for this reason, it is the most widely adopted technology in consumer devices. However, today most of commercial systems are not ready to be used in a pandemic situation as the COVID-19. The most critical weaknesses are the limited energy supply due to the battery small size and limited computational resources.

%
A new recent trend is to couple low power design with machine learning on microcontrollers (MCUs), energy-efficient wireless communication, and energy harvesting to accurately re-design smartwatches to not only help the management of chronic diseases but to potentially play a key role in providing early infection warnings~\cite{colaneri2020sars}.
This paper presents the design and implementation of a hardware-firmware open-source smartwatch called Health Watch (H-Watch). H-Watch combines multi-sensors for health monitoring, an ARM-Cortex-M4F for data acquisition and processing, wireless communication, and energy harvesting to achieve a long-lasting intelligent device. The main H-Watch features are low-power in the range of few mW peak, sub mW average, the capability to run artificial neural networks on-board, a multi-source energy harvesting, high integration resulting in a small form factor, and its novel 5G Narrow Band Internet of Things (NB-IoT) communication. Due to the ultra-low power design and the aggressive power management, H-Watch can automatically and continuously measure dissolved oxygen, heart rate, temperature, respiration rate, motion, and audio signals requiring an average power of 5.9~mW. Thanks to the advanced low-power design and photovoltaic energy harvesting, the device lifetime reaches up to 10 working days with \SI{500}{\lux} of average light. Anyway, it lasts 9 days with a single battery charge. 

H-Watch supports a wide range of connectivity options that can also be used to track and exchange alarms and data with local devices, for example, a smartphone, through the BLE 5.0 connection. Direct global connectivity is also available, using NB-IoT for tracking. NB-IoT is a novel protocol standardized by 3GPP. It is also known as LTE Cat-NB1(NB2) and belongs to Low Power Wide Area Network (LPWAN) technologies that could work virtually anywhere if the 4G (or 5G) infrastructure is present. It can send alarms and data directly to secure servers, such as time traces from on-board sensors. 

To allow researchers and engineers to exploit the features and the hardware-firmware co-design of H-Watch, this paper open-sources\footnote{\url{https://github.com/ETH-PBL/H-Watch}} all the hardware schematics and the layout as well as software libraries for sensors and peripherals, providing a platform to analyze, classify and study a broad infected population sample helping for remote health assistance and diagnosis. Moreover, the repository will provide a library and tools to run Artificial Intelligence (AI) algorithms for on-board data analysis. Indeed, the H-Watch aims to run pre-trained AI models, such as neural networks, for in-situ feature extraction and classification. In this work, we also provide system energy consumption to estimate the battery life with and without the energy harvester. 

%

%
\section{System Architecture}
\begin{figure}[t]
\centerline{\includegraphics[width=1\columnwidth]{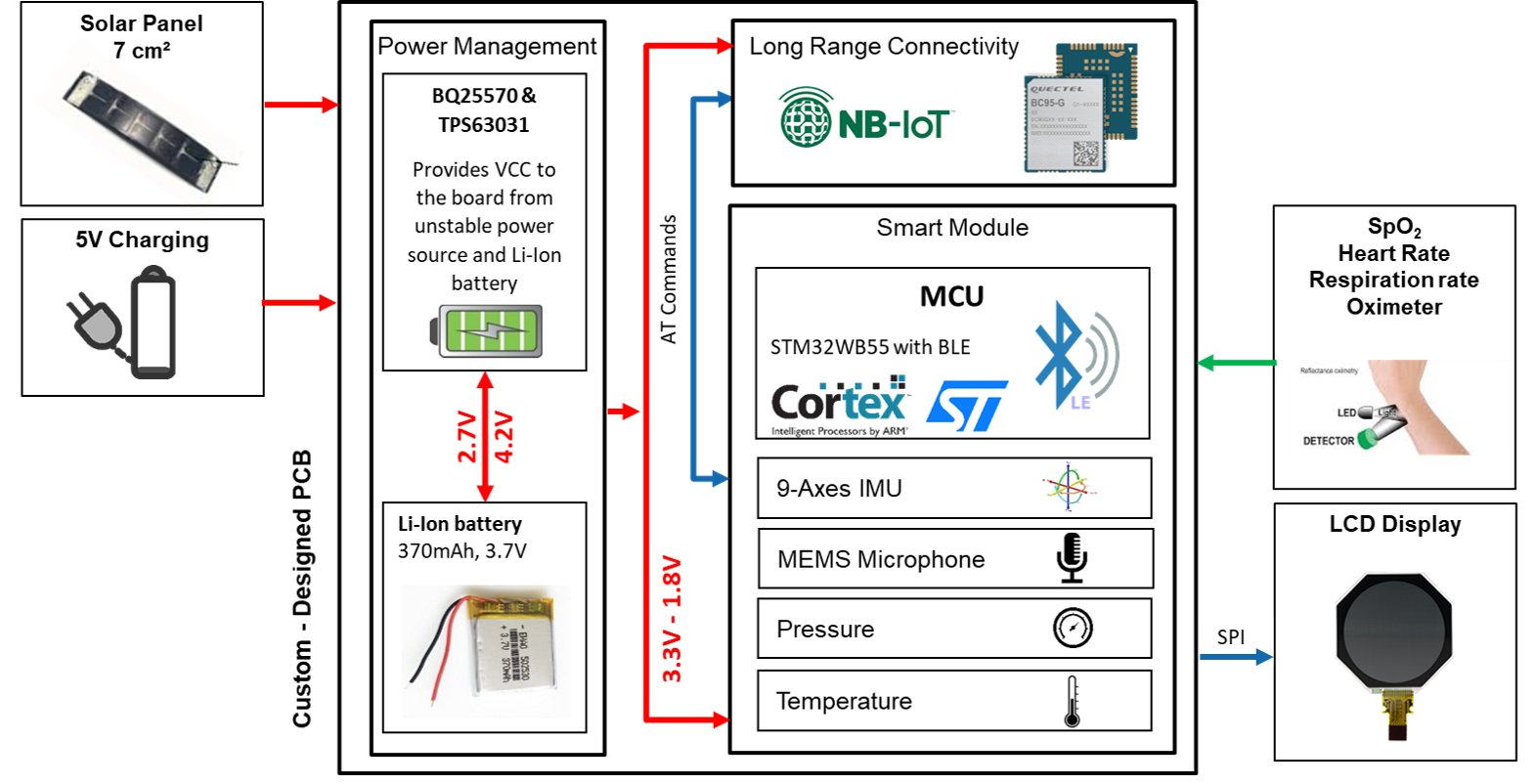}}
\caption{H-Watch logic schematic}
\label{fig:hwatch}
\end{figure}
Figure~\ref{fig:hwatch} shows the H-Watch architecture. From left to right, the figure illustrates the power management (PM), which is specifically designed to achieve extremely low power consumption and to handle both a Li-Ion 370~mAh battery and a 7 $cm^2$ solar panel. The sub-system includes the STM32WB55RGV6 SoC (STM32 hereafter) from ST Microelectronics that manages the sensor acquisition and the wireless connectivity. In parallel with four internal sensors, namely 6-axes IMU (LSM6DSM), 6-axes magnetometer (LSM303AGR), a MEMS Microphone (MP34DT05TR), skin temperature and pressure sensors (LPS22HB), the H-Watch features a low power LCD display (LS012B7DH02). In addition, it features the MAX30101, an integrated SoC for pulse oximetry and heart rate monitoring at low power consumption, 5.5~mW.  H-Watch can communicate with local devices, such as smartphones and laptops, and directly to the cloud through an embedded NB-IoT transceiver (Quectel BC95-G) and the BLE interface.
%
%

\subsection{Power supply and Energy Harvesting}
\begin{figure}[t]
\centerline{\includegraphics[width=0.9\columnwidth]{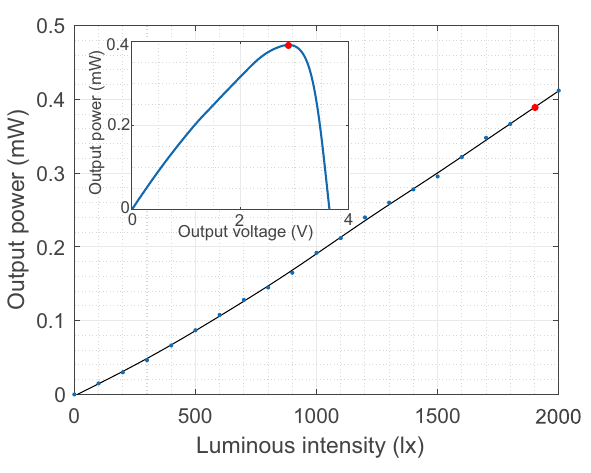}}
\caption{Solar cell output power in matched condition. The inlet shows the influence of the transducer load on the output power in a single harvesting point at \SI{1900}{\lux}.}
\label{fig:eh}
\end{figure}

The power supply sub-system is designed around the BQ25570 and the TPS63031 from Texas Instruments.
BQ25570 is today the most efficient energy harvester chip on the market. It is used to recharge the Li-Ion battery with the flexible solar panel wrapped around the wrist. The BQ25570 achieves high conversion efficiency (90\%) as it periodically adjusts its internal input-impedance to handle Maximum Power Point (MPP) of the energy source; indeed, this parameter changes with the illumination environment.
The BQ25570 has the function of charging the batteries using an integrated boost-converter, and, simultaneously, TPS63031 supplies the system exploiting an integrated high-efficiency buck-boost converter (Table~\ref{cpuemodel} - $V_{DD}$), specifically selected to compensate input voltage oscillations coming from the discontinuous energy source.
The whole supply voltage has been chosen to be 3.3~V, providing a single supply cluster for ICs. This was a specific design choice to decrease the PCB size, as fewer external components are required. Furthermore, it is the lowest voltage supported by the majority of ICs used in the circuit, minimizing the system energy consumption. 
From a wearable application point of view, the effective light spectrum, the size, the flexibility of the cells, and output power, are essential requirements. 
We selected a flexible panel, SP3-12 by Flexsolarcells\footnote{http://www.flexexsolarcells.com}, that measures only $7~cm^2$, which is directly used as H-Watch wrist band.
The energy harvesting capability of the small-sized cell is shown in Figure~\ref{fig:eh}. For the measurement, a Roline RO1332 lux meter has been placed with the solar cell in a darkened chamber artificially illuminated by a controlled broadband light source. The cell output power in matched conditions is measured for illumination from 0 to \SI{2000}{\lux}, which corresponds with the recommended lighting conditions for offices or workspaces. The importance of maximum power point tracking is highlighted in the inset of Figure~\ref{fig:eh} measured at the static illumination of \SI{1900}{\lux}.

In a typical indoor scenario of \SI{500}{\lux} with artificial lighting, the circuit allows to harvest \SI{73}{\micro\watt} with a conversion efficiency of \SI{92}{\%} and a storage voltage of \SI{3.7}{\volt}. In outdoor light conditions of \SI{10}{\kilo\lux}  more than \SI{15}{\milli\watt} can be achieved. 
The power generation of each energy source and the power consumption of each internal sub-system (Table~\ref{cpuemodel}) have been evaluated separately under controlled conditions. We measured the power intake through the Keysight B2900A Series Precision Source/Measure Unit (SMU). 

\subsection{Smart Module  and on-board Processing}
As demonstrated in previous works~\cite{magno2019self}, the IMU can be beneficial to improve the quality of the measurement when the user is moving~\cite{mohan2016measurement}. The pulse oximeter non-invasively measures the blood oxygenation by shining light at two different wavelengths into the wrist and by analyzing the pulsatile component of the reflected
signal~\cite{singh2017infant}. The oximeter sensor controls two LEDs and converts the analog signal from its photodiode into a digital representation for the connected microcontroller. The case form ensures adequate contact with the skin.

The on-board processing capability consists of the STM32 that integrates multiple hardware accelerators (data processing, controlling sensing, communication) in a miniaturized ($8~mm~\times~8~mm$) single IC. The microcontroller has a dual-core architecture with an ARM Cortex M4F for processing and an ARM Cortex M0+ dedicated to the Bluetooth stack.
The power-optimized ARM Cortex M0+ runs autonomously, allowing the rest of the processor to stay in standby mode. Its radio controller is a further reason for selecting this MCU, the patient contact tracing from the BLE advertising mode can be acquired and stored in memory without involving the ARM Cortex M4, improving the energy efficiency of the whole system.  
On-board classification and feature extraction are available through two different tools: X-CUBE-AI from ST Microelectronics is an expansion package that extends the standard STM32 capabilities with automatic conversion of pre-trained Neural Network and ANSI C code generation, and FANN-ON-MCU, ARM Cortex-M, and low power micro-controllers~\cite{wang2020fann}. The latter is a free open source neural network library, which implements instruction optimized (exploiting ARM CMSIS-NN) artificial neural networks in fixed and floating-point computations. It supports C code for both fully connected and sparsely connected networks.
In the GitHub repository\footnote{\url{https://github.com/ETH-PBL/H-Watch}} we release two machine learning examples using both tools. Classification and feature extraction go beyond this paper's scope, which focuses on the hardware and energy profile description.
From our experience, temporal convolutional networks, based on 1D-Convolutional layers, and more widely used MLP feed-forward fully connected networks, need an execution time between 21~ms and 500~ms on the STM32 @ 64~MHz. These values are defined by sampling and classifying heart rate, SpO2, and accelerometer data to extract the patient's health condition.
\subsection{Wireless Connectivity}
H-Watch hosts a dual-radio sub-system, including BLE 5.0 and NB-IoT. In particular, NB-IoT can stream at 170~Kbps with 164~dB of link budget, enabling a wide signal coverage. Despite the long-range communication capabilities, the transceiver needs just 31~uJ per bit when transmitting~\cite{ballerini2020nb}.
H-Watch supports BLE 5.0 for privacy-preserving contact tracing~\cite{zens2020app}, to be compliant with standard protocols used by smartphone Apps, and the NB-IoT for optional geo-tracking and bridge-less communication to secure remote servers, to stream results from on-board processing. 
However, the NB-IoT features a non-negligible energy overhead due to the cellular complex infrastructure. NB-IoT has the dual advantage of communicating through the cellular network like regular smartphones and transferring sensitive data by connecting directly to secure servers, even for low-cost miniaturized devices without the need to go through complex software stacks.
In~\cite{ballerini2020nb}, authors extensively studied and characterized the NB-IoT protocol, where the energy is lightly dependent on the packed size, and heavily affected by the signal strength indicator (RSSI). 
The average energy per packet is consequently modeled in Table~\ref{cpuemodel}, where three different $I^{BC95G}_{active}$ coverage conditions are provided. 

\section{Experimental Results}
\begin{figure}[t]
\centerline{\includegraphics[width=0.9\columnwidth]{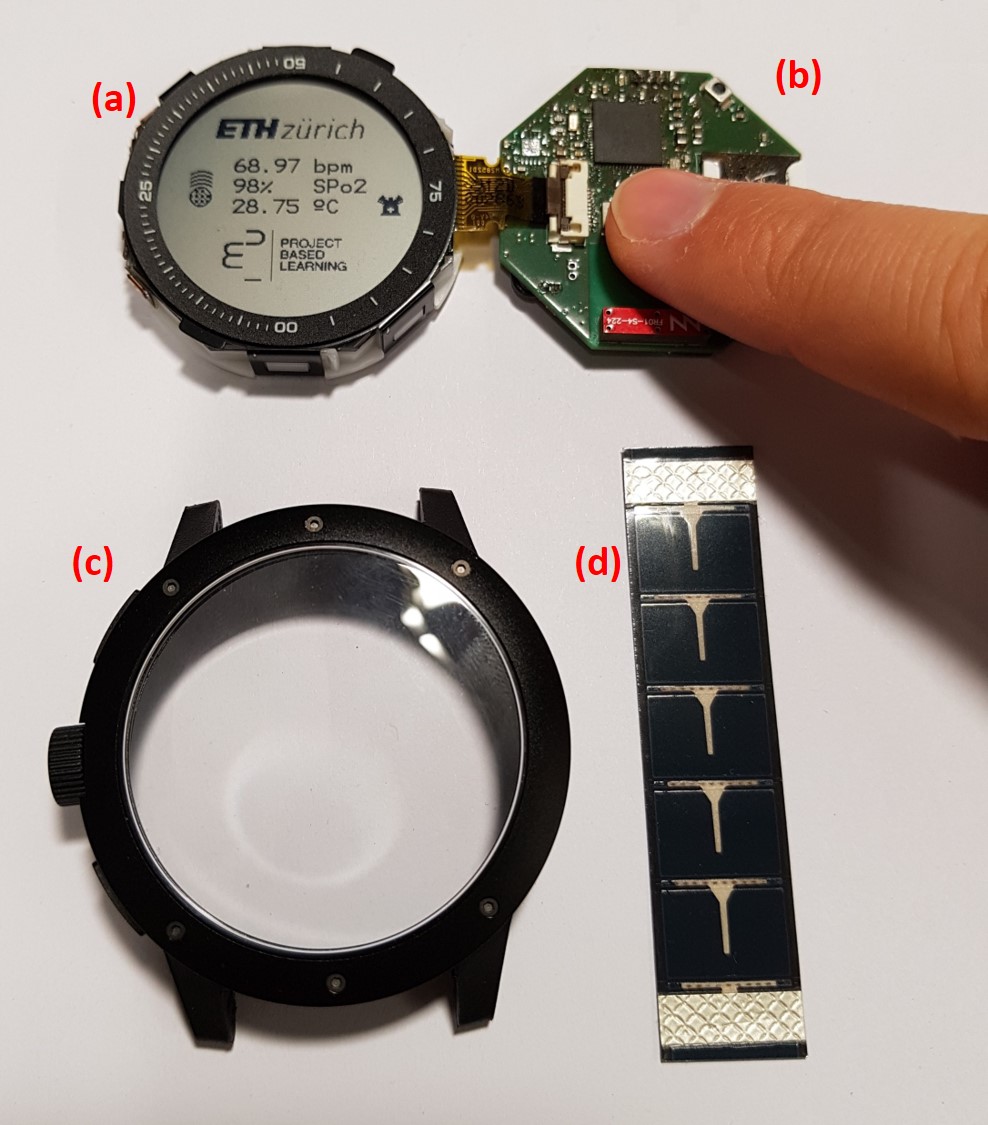}}
\caption{H-Watch mechanical cross-section. (a) LS012B7DH02 LCD display; (b) H-Watch's PCB, (c) Watch case, (d) SP3-12 flexible solar panel.}
\label{fig:hwatchpic}
\end{figure}

 H-Watch has been developed to evaluate functionalities and performances in terms of low power and lifetime.  Figure~\ref{fig:hwatchpic} shows the H-Watch mechanical cross-section in which it presents the flexible solar panel, PCB board, the display, and the clock case. The proposed solution is fully wearable, plug \& play, and it is comfortable to wear, allowing long term measurements without annoying the observed patient.
Table~\ref{cpuemodel} collects and presents the sub-system power consumption for each H-Watch ICs. In this paper, we consider four operation modes. Sleep mode in which all the sensors and radios are off, but the real-time clock and the display are active to enable periodic wake-up, it needs $97~\mu W$. Advertising mode, it is similar to Sleep mode, but the BLE is advertising at 1~Hz and 0~dBm. It is dedicated to contact tracing and consumes only $226~\mu W$. In motion detection mode, with the accelerometer and skin temperature enabled, the H-Watch needs 1.75~mW on average for collecting and processing the data. Lastly, with the full operation mode, in which the watch performs human activity and health classification (oximeter and Heart-rate enabled), the power consumption increases up to 10~mW. These results consider the DC/DC efficiency, named $V_{DD}$ and $V^{MAX30101}_{DD}$ in Table~\ref{cpuemodel}.
In full operation mode, the battery can support the H-Watch for 5 days, while in advertising mode, it exceeds 1 month of operations. Duty cycling between full and motion at 50\%, the H-Watch lifetime is 9 days. One NB-IoT packet per day is considered in these conditions, which is used to send highly compressed information to secure servers after local processing.
In the worst case, when the application layer requires long-range connectivity, the battery longevity is heavily affected. With $RSSI > -95~dBm$, each uplink needs approximately the equivalent amount of energy to run the full operation mode for 1 minute, and with $RSSI < -110~dBm $ it grows up to 6 minutes. With good coverage ($RSSI > -95~dBm$), sending 2.2~MB of data requires approximately 50\% of the battery capacity, while, in the worst case, the uplink volume decreases to 0.6~MB. It is clear that the NB-IoT supports only few packets per day, mainly streaming alarms or pre-extracted features from on-board processing. 
With the help of the energy harvester, H-Watch extends the monitoring time.
Considering an indoor scenario with \SI{500}{\lux}, the captured power support sleep and advertising modes, whose life can reach up to two months (8 hours at \SI{500}{\lux} - 16 hours no energy). However, considering outdoor activities, sleep and advertising mode are fully covered by solar panel, while the 50\% duty-cycled mode between full and motion reaches up to 20 days (6 hours at \SI{10}{\kilo\lux} and 4 hours at \SI{500}{\lux}, 14 hours no energy). 
\begin{table}[!t]
    \renewcommand{\arraystretch}{1.3}
    \caption{H-Watch: power and energy profile}
    \centering
    \label{cpuemodel}
    \resizebox{\columnwidth}{!}{
        \begin{tabular}{l l l}
            \hline\hline \\[-3mm]
            \multicolumn{1}{c}{Symbol} & \multicolumn{1}{c}{Description} & \multicolumn{1}{c}{\pbox{20cm}{Value}}  \\[1.6ex] \hline
            $V_{DD}$ & 3.3~V & Eff. 90\% \\
            $V^{MAX30101}_{DD}$ & 5~V & Eff. 80\% \\
            \hline
            $I_{active}^{STM32}$  &  7.59~mA & 25~mW \\ 
            $I_{idle}^{STM32}$ & 4.15~mA & 13.7~mW \\
            $I_{stop}^{STM32}$ & $2.45~\mu$A & $8.1~\mu W$ \\
            $I_{BLE}^{STM32*}$ & $30~\mu A$ & $99~\mu W$\\
            \hline
            $I^{MAX30101}_{active}$ & $1100~\mu A$ & $5.5~mW$ \\
            $I^{MAX30101}_{shutdown}$ & $0.7~\mu A$ & $3.5~\mu W$ \\
            $I^{LSM303}_{active}$ & $204~\mu A$ & $673~\mu W$ \\
            $I^{LSM303}_{shutdown}$ & $2~\mu A$ & $6.6~\mu W$ \\
            $I^{LPS22HB}_{active}$ & $12~\mu A$ & $40~\mu W$ \\
            $I^{LPS22HB}_{shutdown}$ & $1~\mu A$ & $3.3~\mu W$ \\
            $I^{LSM6DS}_{active}$ & $9~\mu A$ & $30~\mu W$ \\
            $I^{LSM6DS}_{shutdown}$ & $3~\mu A$ & $10~\mu W$ \\
            $I^{MP34DT}_{active}$ & $30~\mu A$ & $99~\mu W$ \\
            $I^{MP34DT}_{shutdown}$ & $1~\mu A$ & $3.3~\mu W$ \\
            \hline
            $I^{BC95G}_{active \bowtie}$ & $RSSI > -95~dBm$ & $1.077~J$ \\
            $I^{BC95G}_{active \bowtie}$ & $-95~dBm > RSSI > -110~dBm $ & $1.422~J$ \\
            $I^{BC95G}_{active \bowtie}$ & $RSSI < -110~dBm $ & $4.071~J$ \\
            $I^{BC95G}_{PSM}$ & $4~\mu A$ & $13.2~\mu W$ \\
            \hline
            $I^{DISPLAY}_{update}$ & $15.2~\mu A$ & $50~\mu W$ \\
            $I^{DISPLAY}_{static}$ & $12.1~\mu A$ & $40~\mu W$ \\
            \hline\hline
            \multicolumn{3}{l}{$^*$ STM32 BLE current advertising (0~dBm; 1~s; 31~B).} \\
            \multicolumn{3}{l}{$^{\bowtie}$ Mean energy per uplink.} 
        \end{tabular}
    }
\end{table}

\section{Conclusions}
This paper proposed the design and implementation of an open-source wearable long-lasting smart monitoring platform for health monitoring and tracking, which can provide direct cloud connectivity through state-of-the-art NB-IoT cellular technology. 
The H-Watch is based on widely available off-the-shelf components; however, it is designed with low power and on-board intelligence in mind. By continuously measuring the blood oxygenation and heart rate with a sampling rate of 50 Hz, accurate results can be achieved with a battery life of 9 days or 20 days, respectively non-using and using the solar energy harvester. 
%
%

%
Finally, H-Wach is also a low-cost platform. The MCU and the energy harvester cost approximatively 10\euro~1KQty, while the BC95-G, the NB-IoT radio transceiver, is the most expensive part, 15\euro. Both external sensors are commonly used in wearable devices, which cost just 2\euro~each. The total cost of off-the-shelf electronics and battery is below 40\euro. The solar panel costs 4\euro~and the battery 2\euro, while the display is included in the watch frame: 35\euro. 
Future works will focus on the sensor data acquisition to verify the accuracy versus a medical device and, on machine learning, to validate early detection. 
\section*{Acknowledgment}
The authors recognise and thank the European Open Science Cloud (EOSC) for financial support under COVID-19 research to open access grant agreements, program call H2020-INFRAEOSC-2018-4 grant number 831644.

\newpage

\bibliographystyle{unsrt}
\bibliography{main}

\end{document}